\documentclass{aa}  
\usepackage{xcolor}
\usepackage{graphicx}
\usepackage{txfonts}
\usepackage{hyperref}
\usepackage{orcidlink}
\begin{document}

   \title{High-resolution radio imaging of TGSS\,J1530$+$1049, a radio galaxy in a dense environment at $z=4$}
\titlerunning{High-resolution radio imaging of TGSS\,J1530$+$1049}

   \author{K. \'E. Gab\'anyi
          \inst{\ref{inst1},\ref{inst2},\ref{inst3}, \ref{inst4}}\orcidlink{0000-0003-1020-1597}
          \and
          S. Frey
          \inst{\ref{inst2},\ref{inst5}}\orcidlink{0000-0003-3079-1889}
          \and
          L. I. Gurvits\inst{\ref{inst6},\ref{inst7}}\orcidlink{0000-0002-0694-2459} 
          \and
          Z. Paragi\inst{\ref{inst6}}\orcidlink{0000-0002-5195-335X}
          \and
          K. Perger\inst{\ref{inst2}}\orcidlink{0000-0002-6044-6069}
          \and
          A. Saxena\inst{\ref{inst8},\ref{inst9}}\orcidlink{0000-0001-5333-9970}
          \and 
          \\ R. A. Overzier\inst{\ref{inst10},\ref{inst11},\ref{inst12}}\orcidlink{0000-0002-8214-7617}
          \and
          M.
          Villar-Mart\'{\i}n\inst{\ref{inst13}}
          \and
          V. Reynaldi\inst{\ref{inst14}, \ref{inst15}}\orcidlink{0000-0002-6472-6711}
          \and
          G. Miley\inst{\ref{inst10}}
          \and
          H. J. A. R\"ottgering\inst{\ref{inst10}}
          \and
          A. Humphrey\inst{\ref{inst16}, \ref{inst17}}
          \and
          Gy. Mez\H{o}\inst{\ref{inst2}}\orcidlink{0000-0002-0686-7479}
}
   \institute{Department of  Astronomy, Institute of Physics and Astronomy, ELTE E\"otv\"os Lor\'and University,
              P\'azm\'any P\'eter s\'et\'any 1/A, 1117 Budapest, Hungary\\
              \email{k.gabanyi@astro.elte.hu} \label{inst1}
         \and
             Konkoly Observatory, HUN-REN Research Centre for Astronomy and Earth Sciences, MTA Centre of Excellence, Konkoly-Thege Mikl\'os \'ut 15-17, 1121 Budapest, Hungary \label{inst2}
        \and
             HUN-REN--ELTE Extragalactic Astrophysics Research Group, ELTE E\"otv\"os Lor\'and University,
              P\'azm\'any P\'eter s\'et\'any 1/A, 1117 Budapest, Hungary
              \label{inst3}
       \and
       Institute of Astronomy, Faculty of Physics, Astronomy and Informatics, Nicolaus Copernicus University, Grudzi\c adzka 5, 87-100 Toru\'n, Poland
       \label{inst4}
        \and
            Institute of Physics and Astronomy, ELTE E\"otv\"os Lor\'and University,
              P\'azm\'any P\'eter s\'et\'any 1/A, 1117 Budapest, Hungary \label{inst5}
        \and
        Joint Institute for VLBI ERIC, Oude Hoogeveensedijk 4, 7991 PD Dwingeloo, The Netherlands \label{inst6}
        \and
        Faculty of Aerospace Engineering, Delft University of Technology, Kluyverweg 1, 2629 HS, Delft, The Netherlands \label{inst7}
        \and
        Department of Physics, University of Oxford, Denys Wilkinson Building, Keble Road, Oxford, OX1 3RH, UK 
        \label{inst8}
        \and
        Department of Physics and Astronomy, University College London, Gower Street, London WC1E 6BT, UK \label{inst9} 
        \and
        Leiden Observatory, University of Leiden, Niels Bohrweg 2, 2333 CA Leiden, The Netherlands \label{inst10}
        \and
        Observat\'orio Nacional/MCTI, Rua General Jos\'e Cristino, 77, Sao Crist\'ovao, Rio de Janeiro, RJ 20921 400, Brazil \label{inst11}
        \and
        TNO, Oude Waalsdorperweg 63, 2597 AK Den Haag, The Netherlands \label{inst12}
        \and
        Centro de Astrobiolog\'{\i}a (CAB), CSIC-INTA, Ctra. de Ajalvir, km 4, E-28850 Torrej\'on de Ardoz, Madrid, Spain
        \label{inst13}
        \and
        Instituto de Astrof\'{\i}sica de La Plata, CONICET -- UNLP, Paseo del Bosque, B1900FWA La Plata, Argentina
        \label{inst14}
        \and
        Facultad de Ciencas Astron\'omicas y Geof\'{\i}sicas, Universidad Nacional de La Plata, B19000FWA La Plata, Argentina
        \label{inst15}
        \and
        DTx -- Digital Transformation CoLab, Building 1, Azur\'em Campus, University of Minho, 4800-058 Guimar\~aes, Portugal \label{inst16}
        \and
        Instituto de Astrof\'isica e Ci\^encias do Espa\c{c}o, Universidade do Porto, CAUP, Rua das Estrelas, PT4150-762 Porto, Portugal \label{inst17}
       }

   \date{Received xx, 2025; accepted }

  \abstract
   {High-redshift radio galaxies can provide important insights into the structure formation and galaxy evolution at earlier cosmological epochs. TGSS\,J1530$+$1049 was selected as a candidate high-redshift radio galaxy, based on its very steep radio spectrum. Subsequent observations with the {\it James Webb Space Telescope} ({\it JWST}) presented in a companion paper have shown that it is located at a redshift $z=4.0$. The {\it JWST} data furthermore showed that the radio source is part of one of the densest structures of galaxies and ionized gas known at these redshifts. The complex system qualitatively resembles a massive (cluster) galaxy forming early through a rapid succession of mergers.}
   {TGSS\,J1530$+$1049 is an unresolved source down to $\sim 0.6\arcsec$ scale in multiple radio surveys. To reveal its high-resolution radio structure and allow for a detailed comparison with JWST observations, we studied its morphology at various angular scales with different radio interferometric instruments.}
   {We observed TGSS\,J1530$+$1049 at milliarcsecond (mas) scale angular resolution with the European VLBI Network (EVN), and at $\sim 100$-mas scale resolution with the enhanced Multi-Element Remotely Linked Interferometer Network (e-MERLIN).}
   {We recovered a complex north--south oriented structure with steep-spectrum radio-emitting features, which are associated with lobes and hot spots of a jetted active galactic nucleus. However, the centre of the radio galaxy proved to be too faint at cm wavelengths to be unambiguously detected in our observations. Nevertheless, the linear size ($\sim5.5$ kpc) and the radio power ($L_\mathrm{1.4GHz}\approx3\times10^{27}$ W Hz$^{-1}$) place it among the so-called medium-sized symmetric objects, a smaller and/or confined version of larger radio galaxies. Comparison between the radio morphology and that of the ionized gas as observed with the NIRSpec IFU on {\it JWST} shows that the two are closely aligned. However, the optical emission line gas extends out to $\sim25$ kpc, which is well beyond the detected radio structures.}
   {}


   \maketitle
%

\section{Introduction}

High-redshift ($z\gtrsim4$) active galactic nuclei (AGN) are essential in studying the growth of supermassive black holes (SMBH) and the galaxy evolution in the early Universe. The radio-emitting subsample of these sources is of great importance since they can be studied with the highest angular resolution provided by radio interferometric technique. Among them, high-redshift radio galaxies are important constituents to study the evolution of large-scale structure in the Universe since they are often found in the centre of clusters and proto-clusters, and can be the progenitors of massive galaxies. At early cosmological epochs, the radio galaxies are expected to be young, compact systems.

Historically, it was proposed that compact radio sources with ultra-steep radio spectra could be good candidates for high-redshift radio galaxies \citep[e.g.,][]{Blumenthal_Miley, Rottgering1994, Chambers_1996}. Various authors use different definitions for the ultra-steep-spectrum (USS) sources (see, e.g., \citet{Brude+1995Ap&SS}, and, for a later summary, \citet{Coppejans2017}), but more recently \cite{Saxena_sample} collected a sample of USS sources with radio spectral indices $\alpha<-1.3$ ($\alpha$ is defined as $S \propto \nu^\alpha$, where $S$ is the flux density and $\nu$ the observing frequency) measured between $150$\,MHz and $1.4$\,GHz. To enable further efficient selection of potential high-redshift sources, \citet{Saxena_sample} focused on those which remained undetected in optical and infrared surveys. 

TGSS\,J1530$+$1049 (hereafter J1530$+$1049) was one of the brightest objects in their sample. In a follow-up radio observation of the sample, conducted with the Karl G. Jansky Very Large Array (VLA) at $1.4$\,GHz in its most extended A configuration, J1530$+$1049 was detected as a source well-described by a single Gaussian brightness distribution with a deconvolved size of $\sim 0.6\arcsec$, and a flux density of $(7.5\pm0.1)$\,mJy. Optical spectroscopic observation at the radio position revealed a single emission line that was erroneously identified as a Lyman-$\alpha$ line at $z=5.72$ \citep{disc}. However, recent {\it James Webb Space Telescope (JWST)} imaging and spectroscopic observations of the object presented in a companion paper \citep{Saxena_2026} have revealed that the radio emission positionally coincides with a highly complex system consisting of several massive galaxies, diffuse emission and ionized gas knots at $z=4.0$. The picture that has emerged is that of a dense core of interacting galaxies including the host galaxy of the radio source, perhaps signaling the formation of a very massive (central cluster) galaxy.
   
We observed J1530$+$1049 with the technique of very long baseline interferometry (VLBI) using the European VLBI Network (EVN) and the enhanced Multi-Element Remotely Linked Interferometer Network (e-MERLIN) to resolve its radio emission. Preliminary analysis of the EVN observation were presented in \cite{rnaas}, which used the redshift estimate available at that time. In Sect.~\ref{obs}, we describe the observations and data reduction. Results are presented in Sect.~\ref{res}, while our findings are discussed in Sect.~\ref{dis}. The paper is concluded by a summary in Sect.~\ref{sum}.

In the following, we assume a $\Lambda$ Cold Dark Matter cosmological model with a Hubble constant of $H_0=67.7 \,\, \mathrm{km\,s}^{-1}\mathrm{\,Mpc}^{-1}$, matter density parameter of $\Omega_\mathrm{m}=0.31$, and dark energy density parameter of $\Omega_\Lambda=0.69$. At $z=4.0$, the angular size of $1$ milliarcsecond (mas) corresponds to $7.1$\,pc linear size, and the luminosity distance of the source is $D_\mathrm{L}=36676.4$\,Mpc \citep{Cosmocalc}.

\section{Observations and data reduction}
\label{obs}

\begin{figure*}
    \centering
    \includegraphics[width=\textwidth, bb=0 100 970 445, clip]{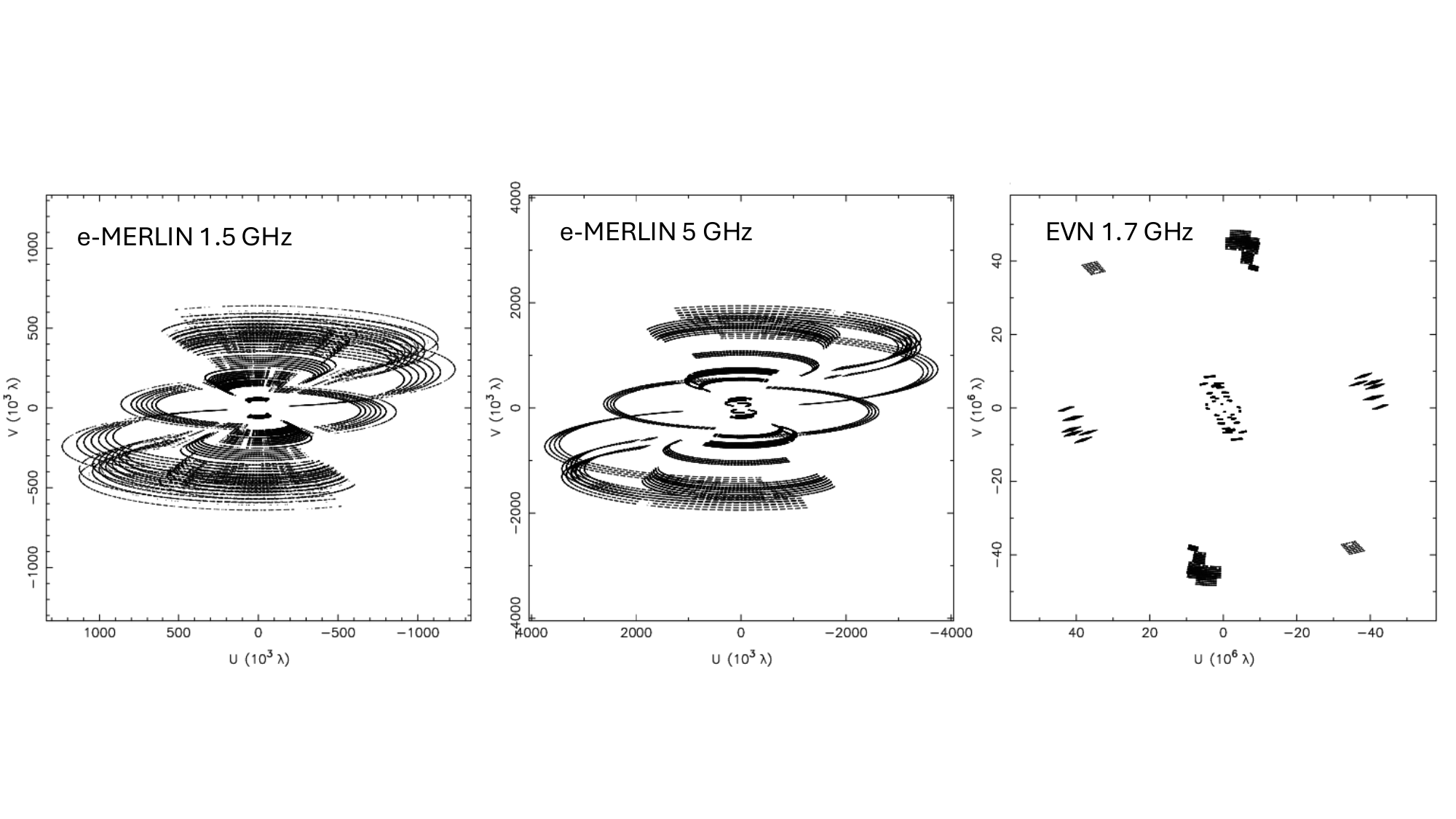}
    \caption{The $(u,v)$-coverages of the three radio interferometric observations of J1530$+$1049, from left to right, the $1.5$-GHz e-MERLIN, the $5$-GHz e-MERLIN, and the $1.7$-GHz EVN observations. We note that the units of the axes are $10^3 \lambda$ for the e-MERLIN observations and $10^6 \lambda$ for the EVN observation, where $\lambda$ indicates the observing wavelength.}
    \label{fig:uvplot}
\end{figure*}

\subsection{EVN observation}
The EVN observation of J1530$+$1049 was conducted on 2018 September 19 at $1.7$\,GHz frequency  (project code: RSG11) in e-VLBI mode \citep{evlbi}. The observing bandwidth was divided into $8$ intermediate frequency channels (IF) with $16$\,MHz bandwidth each. The following antennas provided data: a single antenna of the Westerbork Synthesis Radio Telescope (the Netherlands), Effelsberg (Germany), Medicina (Italy), Onsala (Sweden), Tianma (China), Toru\'n (Poland), Hartebeesthoek (South Africa), and Sardinia (Italy). Sardinia observed in only four IFs. The $(u,v)$-coverage of the observation is shown in the rightmost panel of Fig.\,\ref{fig:uvplot}.

The observation was conducted in phase-referencing mode \citep{p-ref}. The phase-reference calibrator, ICRF\,J152502.9$+$110744 (hereafter J1525$+$1107), and the target source were observed alternately, in cycles with $1.5$~min time spent on the calibrator and $4.5$~min on the target source. The total on-target integration time was $57.6$~min. 

The data were reduced with the National Radio Astronomy Observatory (NRAO) Astronomical Image Processing System \citep[\texttt{AIPS},][]{aips} following the standard procedure \citep[e.g.,][]{VLBI_reduc}. A priori amplitude calibration was performed using system temperature measurements at the antenna sites. Ionospheric corrections were conducted based on the total electron content maps from global navigation satellite systems data, and parallactic angle correction was also done. Then fringe fitting was performed on the data of the phase-reference calibrator.  

The resulting calibrator source data were imaged in the \texttt{Difmap} software \citep{difmap} via the hybrid mapping method, which involves subsequent usage of the CLEAN algorithm \citep{clean} and phase self-calibration iterations. As a final step of imaging, we performed amplitude self-calibration and obtained overall gain correction factors for the antennas. These gain corrections were then applied to the data within \texttt{AIPS}. Additionally, the resulting image of the phase-reference calibrator was read back into \texttt{AIPS} to be used for improving the fringe fit. The fringe-fit solutions of the phase-reference calibrator were applied to the target source, which was than imaged in \texttt{Difmap}. Due to the weakness of the main target object, no self-calibration was attempted.

\subsection{e-MERLIN observations}
The e-MERLIN observations (project code: CY8205) took place on 2019 January 29 and February 22 at $5$\,GHz, and on 2019 March 7 and 8 at $1.5$\,GHz. At $1.5$\,GHz, $8$ IFs were used, each with a bandwidth of $64$\,MHz and divided into $128$ spectral channels. However, no signal was recorded in IF$7$. At $5$\,GHz, $4$ IFs were used, each with a bandwidth of $128$\,MHz and divided into $128$ channels. At both frequencies, the Jodrell Bank Mk2, Pickmere, Darnhall, Knockin, Defford, and Cambridge antennas participated in the observations. The Lovell Telescope at Jodrell Bank also joined at $5$\,GHz on the second observing day. The $(u,v)$-coverages of the observations are shown in the left and middle panels of Fig.\,\ref{fig:uvplot}.

Beside the target source, the following calibrators were observed: 3C\,84 as a pointing calibrator, 3C\,286 as a flux density calibrator, OQ\,208 as a bandpass calibrator. The phase calibrator was the same radio quasar used in the EVN observation, (J1525$+$1107). The on-target integration time was $\sim9$\,h and $\sim13.2$\,h at $1.5$ and $5$\,GHz, respectively.

The raw visibility data were pre-processed and calibrated with the e-MERLIN pipeline\footnote{version 0.9} using the Common Astronomy Software Applications \citep[\texttt{CASA},][]{casa_old,casa}\footnote{version 5.4}. Imaging was also performed in \texttt{CASA}. We exported the channel-averaged dataset and additionally imaged it in \texttt{Difmap}. The resulting maps agree with each other. Due to the weakness of the source, no self-calibration was performed.

\subsection{{\it JWST} observations}

J1530$+$1049 was observed with the {\it JWST}/NIRSpec in IFU mode on 2023 July 14--15 (PIs: R.A. Overzier and A. Saxena, program GO\,1964). 
The results of that program involving additional observations with the {\it Hubble Space Telescope} and {\it JWST}/NIRCam are discussed in detail in \cite{Saxena_2026}. In this paper, we present a comparison between the high-resolution radio data and a map of the H$\alpha$ emission as measured with the NIRSpec IFU (see Sect. \ref{sec:jwst}). For details on the observations and construction of the H$\alpha$ map, we refer the reader to the companion paper by \citep{Saxena_2026}.  

\section{Results}
\label{res}

\subsection{EVN observation}

\begin{figure*}
    \centering
    \resizebox{\hsize}{!}
            {\includegraphics[width=12cm]{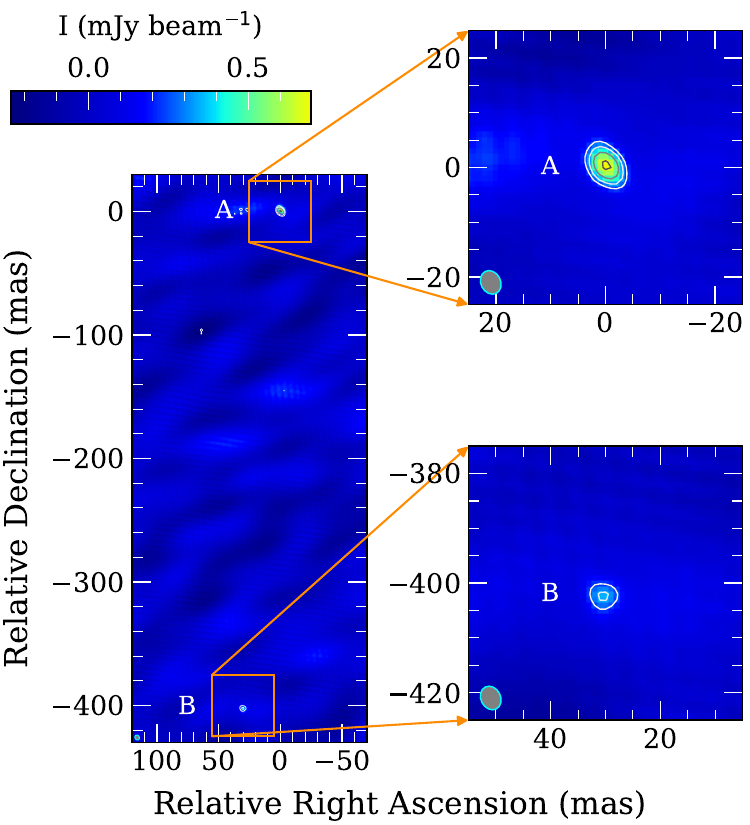}}
      \caption{EVN 1.7-GHz maps of J1530$+$1049. On the right-hand side, the two detected features are shown zoomed-in. The peak intensity of the whole image is $0.70\mathrm{\,mJy\,beam}^{-1}$. The contour levels are at $(\pm0.23, 0.33, 0.47, 0.66)\mathrm{\,mJy\,beam}^{-1}$. The lowest positive contour level corresponds to $5\sigma$ image noise. The restoring beam is shown in the lower left corner of the panels. Its FWHM size is $4.48\mathrm{\,mas} \times 3.61\mathrm{\,mas}$ at a position angle of $27\degr$.
              }
         \label{fig:EVN}
\end{figure*}

Two mas-scale compact features (labeled as A and B) were detected with the EVN at $1.7$\,GHz with a projected separation of $\sim 400$\,mas (Fig.~\ref{fig:EVN}), corresponding to $\sim 2.8$\,kpc at the redshift of the source.

The coordinates of the brighter, northern feature (A) are right ascension $\alpha=15^\mathrm{h} 30^\mathrm{m} 49\fs89029 \pm 0\fs00013$, and declination $\delta=10\degr 49\arcmin 31\farcs1754 \pm 0\farcs002$. When calculating the positional error, we took into account the uncertainty of the coordinates of the phase-reference calibrator source ($0.11$\,mas in right ascension and $0.18$\,mas in declination direction\footnote{According to the latest Radio Fundamental Catalog \citep{rfc} rfc2025b maintained by L. Petrov at \url{https://astrogeo.org/sol/rfc/rfc_2025b/}}), and the error arising from the phase-referencing technique which depends on the target and calibrator source separation, in our case $1\fdg45$. \cite{multiview} estimate this error to be $\sim2$\,mas. \cite{pradel_charlot} obtained a close value of $1.5$\,mas via simulated VLBI observations. We used the larger, more conservative value. It should also be noted that the coordinates of the phase-reference calibrator source were measured at $8.4$\,GHz, while the EVN observation was conducted at $1.7$\,GHz. Frequency-dependent core shift \citep{coreshift} could cause a difference between the International Celestial Reference Frame \citep[ICRF,][]{icrf3} position determined at $8.4$~GHz and the $1.7$-GHz position of the calibrator.

\begin{table}
\caption{Parameters of the Gaussian components fitted to the $1.7$-GHz EVN visibility data of J1530$+$1049. 
}             
\label{tab:evn_modelfit}      
\centering                          
\begin{tabular}{c c c c c }       
\hline\hline                 
ID & $S$ (mJy) & $\theta$ (mas) & $\Delta$RA (mas) & $\Delta$Dec (mas)\\ 
\hline                        
A & $1.96\pm0.30$ & $10.2\pm1.0$ & -- & --\\
B & $0.62\pm0.17$ & $4.0\pm0.8$ & $31.7\pm0.6$ & $-404.8\pm0.6$\\
\hline                                   
\end{tabular}
\newline
Notes: Col.~1 -- component identifier, Col.~2 -- flux density, Col.~3 -- angular size (FWHM), Col.~4 -- relative right ascension, Col.~5 -- relative declination 
\end{table}

To describe the emission quantitatively, we fitted the visibilities with circular Gaussian brightness distribution model components. Since self-calibration cannot be performed, the obtained flux densities are expected to be lower than the real value due to coherence loss \citep{coherence-loss}. Previous works \citep[e.g.][]{mosoni_coherenceloss,gabanyi_coherenceloss} estimated the coherence loss to be $\sim 25$\,\%. The parameters of the two fitted components are given in Table~\ref{tab:evn_modelfit}, where the flux densities have been increased to account for the coherence loss. The errors were calculated according to the formulae given by \cite{error_est}. For the flux density errors, another $10$\,\% was added in quadrature to account for the absolute amplitude calibration inaccuracies. The full width at half-maximum (FWHM) sizes of the fitted Gaussian features, $\sim 10$\,mas and $\sim 4$\,mas, corresponding to $\sim 71$\,pc and $\sim 28$\,pc linear sizes at the source's redshift, quantify their compact nature.

The brightness temperature of the features can be calculated as \citep[e.g.,][]{tb_veres}
\begin{equation}
    T_\mathrm{B}=1.22 \cdot 10^{12} \frac{S}{\nu^2\theta^2} (1+z)\,\,\,\mathrm{K},
\end{equation}
where $S$ is the flux density of the components measured in Jy, $\nu$ the observing frequency in GHz, and $\theta$ the FWHM size in mas. The values obtained are $T_\mathrm{B}^\mathrm{A}=(4.0\pm0.8) \cdot10^{7}$\,K and $T_\mathrm{B}^\mathrm{B}=(8.2\pm3.2) \cdot10^{7}$\,K for components A and B, respectively. Both values well exceed the upper limit given for star-forming galaxies by \cite{condon_radio}, $\sim 10^5$\,K, indicating AGN-related radio emission.

\subsection{e-MERLIN observations}

  \begin{figure}
   \centering
   \includegraphics[width=1\linewidth,  clip]{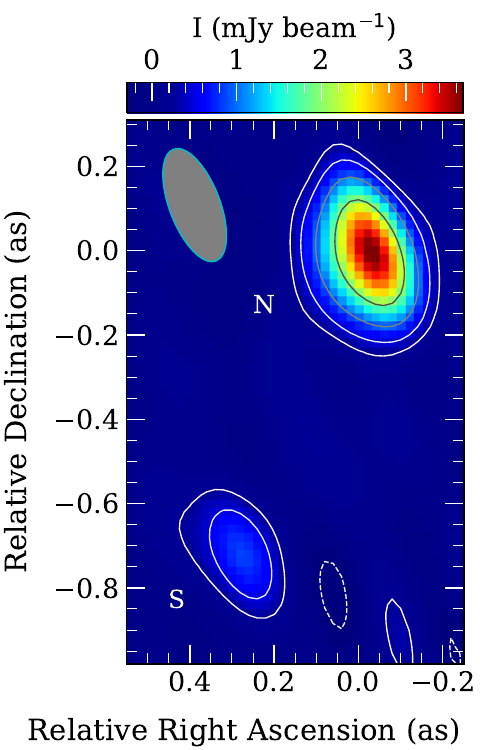}
      \caption{1.5-GHz e-MERLIN image of J1530+1049. The peak intensity is $3.7\mathrm{\,mJy\,beam}^{-1}$. The lowest contours are drawn at $\pm0.24\mathrm{\,mJy\,beam}^{-1}$ corresponding to $6\sigma$ image noise level. Further positive contour levels increase by a factor of two. The restoring beam is shown in the upper left corner, its FWHM size is $284 \mathrm{\,mas} \times 120$\,mas and its major axis is oriented at a position angle of $21\degr$.}
         \label{fig:eMERLIN-L}
   \end{figure}

\begin{figure}
   \centering
   \includegraphics[width=1.\linewidth, clip]{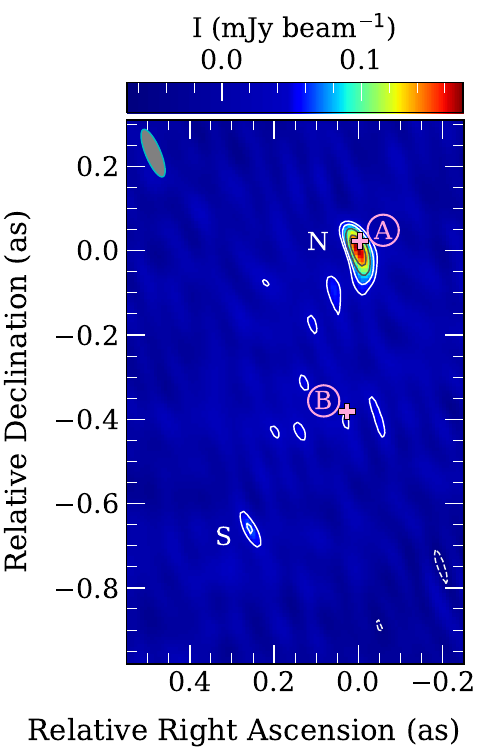}
      \caption{5-GHz e-MERLIN image of J1530+1049. The peak intensity is $0.17\mathrm{\,mJy\,beam}^{-1}$. The lowest contours are drawn at $\pm0.046\mathrm{\,mJy\,beam}^{-1}$ corresponding to $3.5\sigma$ image noise level. Further positive contour levels increase by a factor of $\sqrt{2}$. The restoring beam is shown in the upper left corner, its FWHM size is $121 \mathrm{\,mas} \times 36$\,mas and its major axis is oriented at a position angle of $23\degr$. The pink crosses and letters indicate the positions of the EVN-detected components A (north), and B (south).}
         \label{fig:eMERLIN-C}
   \end{figure}

With e-MERLIN, two features (designated as N and S) could be detected at $1.5$\,GHz with a separation of $\sim780$\,mas (corresponding to $\sim 5.5$\,kpc projected linear size), as shown in Fig.~\ref{fig:eMERLIN-L}. At $5$\,GHz, the brighter northern feature (N) is clearly detected with e-MERLIN. At the approximate distance of feature S, at $\sim 700$\,mas from N, and in the same position angle as seen in Fig.~\ref{fig:eMERLIN-L}, we recover an emission spot at $\sim 5\sigma$ image noise level at $5$\,GHz (Fig.~\ref{fig:eMERLIN-C}).

\begin{table*}
\caption{Parameters of the Gaussian components fitted to the e-MERLIN visibility data of J1530$+$1049. 
}             
\label{tab:modelfit}      
\centering                          
\begin{tabular}{c c c c c c }        
\hline\hline                 
$\nu$ & ID & $S$ & $\theta$ & $\Delta$RA & $\Delta$Dec\\ 
(GHz) & & (mJy) & (mas) & (mas) & (mas) \\
\hline                        
$1.318$ & N & $5.7\pm1.1$ & $86.4\pm2.3$ & -- & --  \\ 
& S & $1.2\pm0.3$ & $116.0\pm19.0$ & $318\pm21$ & $-740\pm 47$ \\
$1.446$ & N & $3.7\pm0.8$ & $77.3\pm8.7$ & -- & -- \\ 
 & S & $0.6\pm0.2$ & $31.7\pm4.3$ & $310\pm19$ & $-710\pm42$\\
$1. 574$ & N & $2.3\pm0.5$ & $73.4\pm3.2$ & -- & -- \\ 
 & S & $0.4\pm0.1$ & $63.5\pm16.9$ & $292\pm17$ & $-742\pm39$ \\
 $1.734$ & N & $3.2\pm0.7$ & $151.6\pm8.1$ & -- & -- \\
\hline
$5.072$ & N & $0.25\pm0.02$ & $37.4\pm 3.1$ & -- & -- \\ 
 & S & $0.08\pm0.03$ & $76.5\pm31.4$ & $247\pm 6$ & $-673\pm 14$ \\
\hline                                   
\end{tabular}
\newline
Notes: Col.~1 -- central frequency, Col.~2 -- component identifier, Col.~3 -- flux density, Col.~4 -- angular size (FWHM), Col.~5 -- relative right ascension, Col.~6 -- relative declination
\end{table*}

We used \texttt{Difmap} to fit the visibilities with circular Gaussian features to quantify the radio emission. Note that the brightness distribution models were fitted directly to the interferometric visibility data and not in the image plane, so the component sizes obtained can be considered as deconvolved sizes. In the $1.5$-GHz band, we averaged the adjacent IFs by two together before fitting the visibilities, to minimize bandwidth smearing effects, but to increase the signal-to-noise ratio. Since IF$7$ did not record data, IF$8$ had to be fitted as it is. In the first three $128$ MHz-wide chunks (i.e. IF pairs), both components, N and S could be fitted in the $1.5$-GHz band. In the last one, where the bandwidth was only $64$\,MHz, the signal-to-noise ratio was not adequate to reliably fit component S.  
In the $5$-GHz band, the components N and S could only be modeled if we used the whole bandwidth for the fit. The parameters of the brighter component N agreed within the errors if the first and second halves of the $5$-GHz band were fitted separately. 

The results of the model fitting are given in Table~\ref{tab:modelfit}. The errors have been estimated using the formulae of \cite{error_est}, and additional $20\%$ and $10\%$ were added quadratically to the flux density errors to account for the amplitude calibration uncertainty of e-MERLIN at $1.5$\,GHz and $5$\,GHz, respectively \citep{Lband_error, Cband_error}. 

\begin{figure}
   \centering
   \includegraphics[width=\hsize, bb=20 40 710 570, clip=]{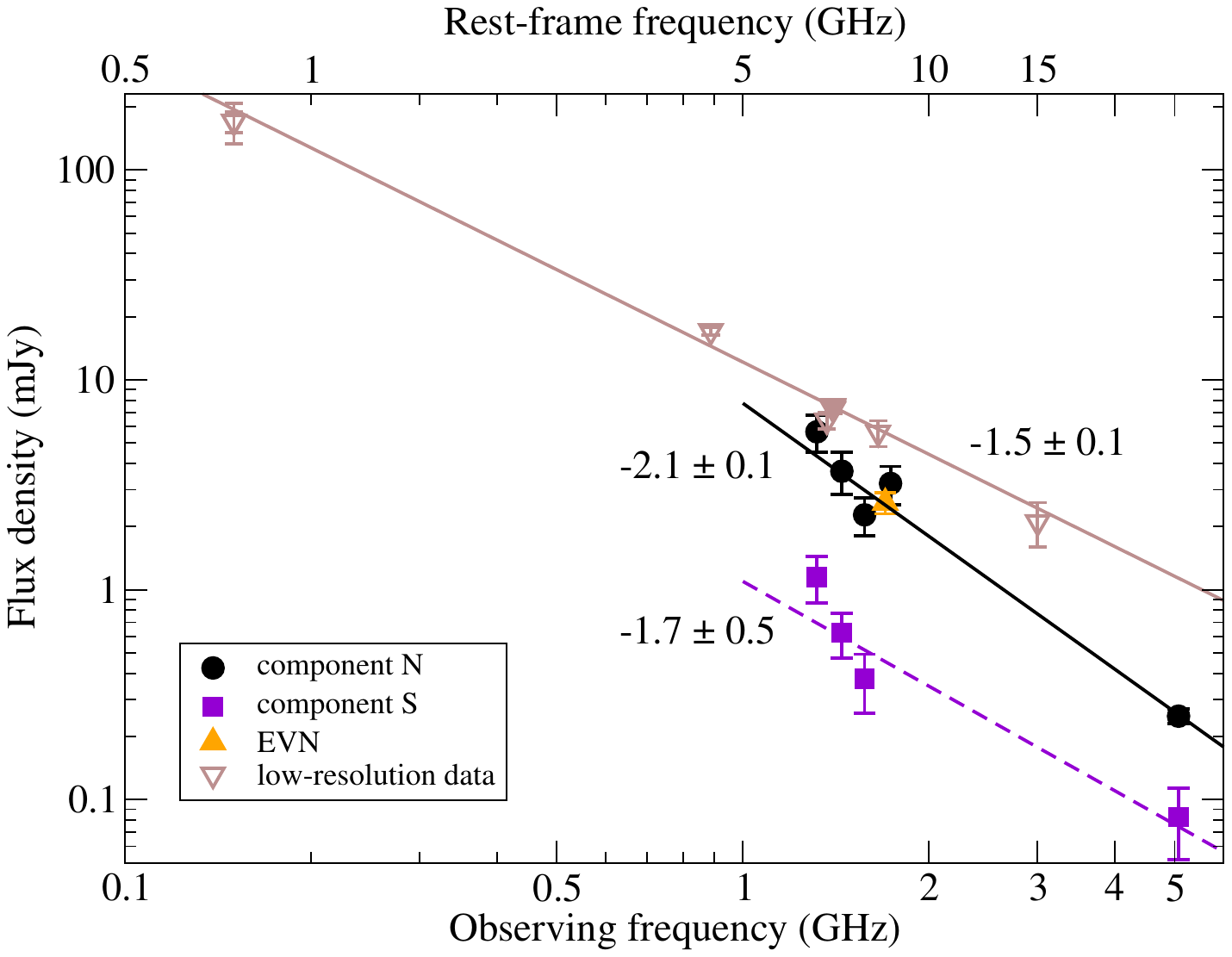}
      \caption{The flux densities of the two Gaussian components that can be fitted to the e-MERLIN visibilities of J1530$+$1049. Circles and squares are for the components N and S, respectively. The solid and dashed lines are power-law fits to the corresponding data points. For comparison, the sum of the flux densities of the EVN-detected components is also shown with a yellow upward triangle. Empty downward brown triangles show lower-resolution flux density measurements from various surveys (see Sect.\,\ref{sec:large_scale} for details), and the brown line indicates the power-law fit to those points. The numbers show the corresponding spectral indices.
              }
         \label{fig:spectralindex}
   \end{figure}

Both components have steep radio spectra (Fig.~\ref{fig:spectralindex}). Their spectral indices are $\alpha_\mathrm{N}=-2.1\pm0.1$ and $\alpha_\mathrm{S}=-1.7\pm0.5$. 

\section{Discussion}
\label{dis}

\subsection{Astrometric registration of the radio components}

\begin{figure*}
    \resizebox{\hsize}{!}{
    \includegraphics[width=12cm, bb=0 100 710 450,clip]{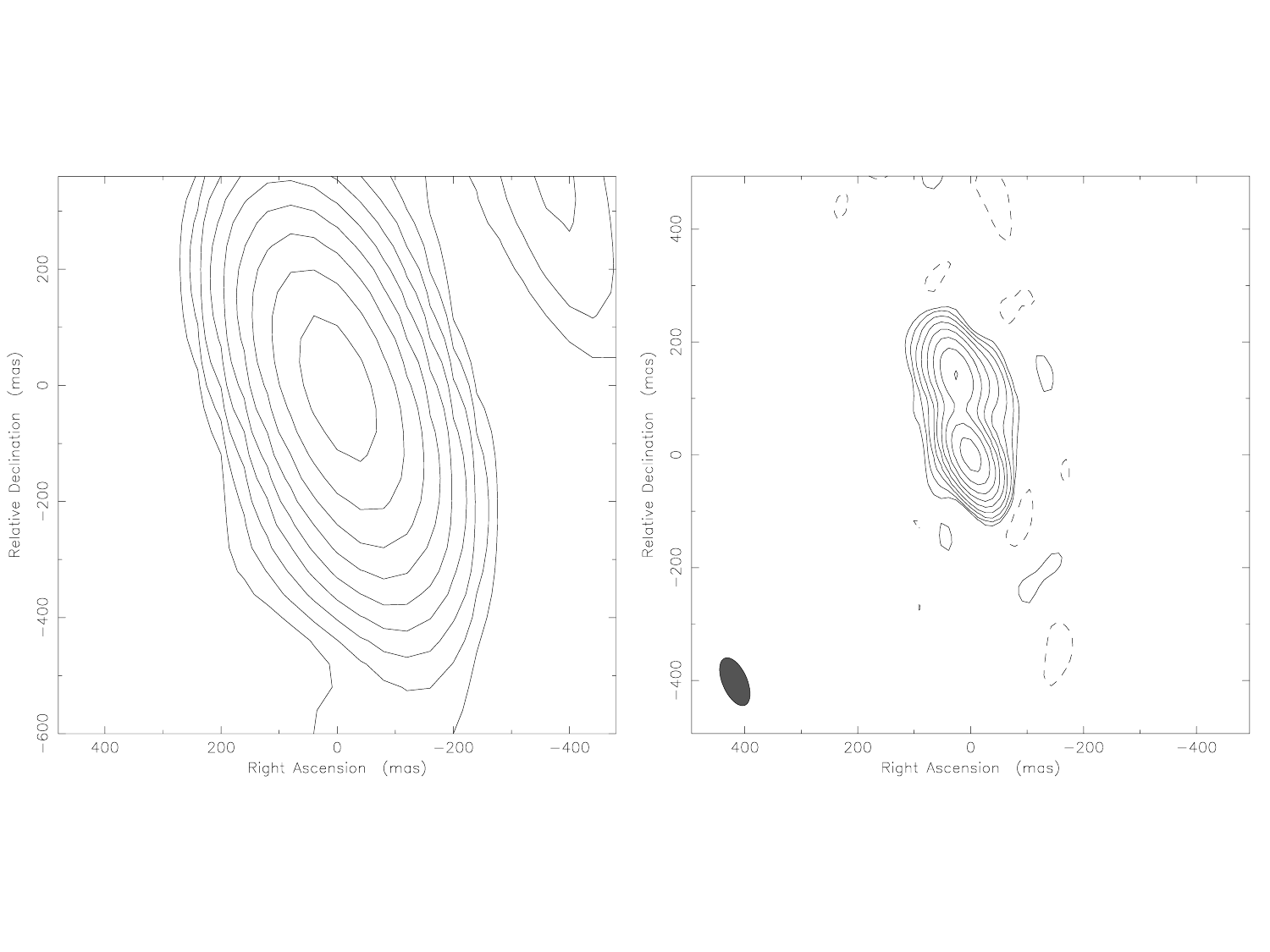}}
    \caption{The e-MERLIN images of the calibrator source, J1525$+$1107. {\it Left:} at $1.5$-GHz. The peak intensity is $420\mathrm{\,mJy\,beam}^{-1}$. The lowest positive contour is drawn at $1.1\mathrm{\,mJy\,beam}^{-1}$ corresponding to $7\sigma$ image noise level. The FWHM size of the restoring beam is $321 \mathrm{\,mas} \times 144$\,mas and its major axis is oriented at a position angle of $21\degr$. {\it Right:} $5$-GHz. The peak intensity is $274\mathrm{\,mJy\,beam}^{-1}$. The lowest positive contour is drawn at $0.75\mathrm{\,mJy\,beam}^{-1}$ corresponding to $7\sigma$ image noise level. The restoring beam is shown in the lower left corner, its FWHM size is $91 \mathrm{\,mas} \times 44$\,mas and its major axis is oriented at a position angle of $24\degr$. The two image cutouts have the same size. To illustrate that the $1.5$-GHz peak corresponds to the secondary rather than the primary brightness peak at $5$~GHz, the images are displaced by $110$~mas with respect to each other in the declination direction.}
    \label{fig:calJ1525}
\end{figure*}

The EVN-detected compact feature A falls within the region of the e-MERLIN-detected brightest feature, N, at $5$\,GHz. However, there is a significant, $\sim110$\,mas offset in the north--south direction between the peak positions measured in the $5$-GHz and $1.5$-GHz e-MERLIN observations. The question arises whether this offset is caused by some physical effect in the target source, J1530$+$1049, or due to a difference in the astrometric registrations of the $1.5$- and $5$-GHz e-MERLIN images. 

The same phase-reference source (J1525$+$1107) was used in all observations. However, it has significant radio emission extended to $\sim100$-mas scale and the position of its brightness peak can be different at different observing frequencies. To illustrate this, the e-MERLIN images of the calibrator are shown in Fig.~\ref{fig:calJ1525}, with a shift of $\sim110$\,mas between the image cutouts to facilitate their comparison. The north--south oriented double feature seen at $5$\,GHz cannot be resolved at $1.5$\,GHz with e-MERLIN. However, if the flux densities of the two components relative to each other change with frequency, i.e. the northern feature has higher or very similar flux density than the southern one at the lower frequency of $1.5$\,GHz, then the position of the brightness peak can shift towards the northern feature which, on the other hand, is relatively fainter at $5$\,GHz. This frequency-dependent change in the location of the brightness peak is further supported by higher-resolution VLBI images of the calibrator taken at $2.3$\,GHz, showing an extended bright feature to the north of the mas-scale compact emitting region (the core) at a distance of $\sim(120-150)$\,mas. These archival images are available at the Astrogeo website\footnote{\url{http://astrogeo.org/cgi-bin/imdb_get_source.csh?source=J1525\%2B1107} maintained by L. Petrov}. Thus we can conclude that the positional mismatch between the $1.5$-GHz and $5$-GHz e-MERLIN images of J1530$+$1049 is caused by different brightness distributions of the phase-reference calibrator source at the two frequencies used (Fig.~\ref{fig:calJ1525}). Therefore, in our target source, the brighter northern features detected by e-MERLIN at the two frequencies (Figs.~\ref{fig:eMERLIN-L} and \ref{fig:eMERLIN-C}) are co-spatial, and the brighter EVN-detected component (A, Fig.~\ref{fig:EVN}) is located within this emitting region. Since the distances and position angles of the target morphology are very similar in the two e-MERLIN images, and no other significant radio emission can be detected in the field at either of the frequencies, we conclude that their cross-identification is solid.

\subsection{The nature of the radio features}

\begin{figure}
    \centering
    \includegraphics[width=\linewidth, bb = 60 10 410 310, clip]{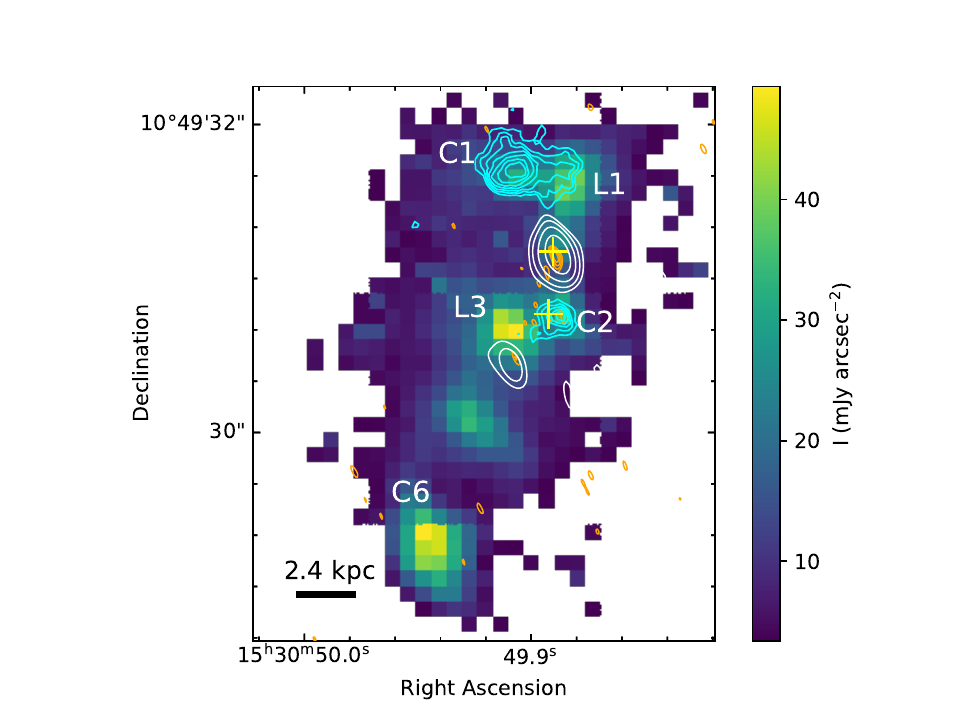}
    \caption{Overlay of the {\it JWST} H$\alpha$+[N\,\textsc{ii}] emission line map, the NIRCam F300M continuum image from \cite{Saxena_2026}, and the e-MERLIN maps of J1530$+$1049. The colour scale shows the continuum-subtracted H$\alpha$ line map, where low-significance pixels have been masked \citep[for details, see ][]{Saxena_2026}. The logarithmically scaled cyan contours show the optical continuum data taken with F300M filter. The lowest contour is drawn at $9.8$\,mJy\,arcsec$^{-2}$. The white and orange contours represent the $1.5$-GHz and the $5$-GHz e-MERLIN data. Radio contour levels are the same as in Figs.\,\ref{fig:eMERLIN-L}, \ref{fig:eMERLIN-C}, except that negative contours are omitted for clarity. Yellow crosses show the positions of the two EVN-detected features (the positional uncertainties are much smaller than their sizes). The C1, C2 and C6 and L1 and L3 labels show the optical continuum and line-emitting regions, respectively as defined in \cite{Saxena_2026}. In the lower left corner the linear scale is displayed.} 
    \label{fig:JWST-eMERLIN}
\end{figure}

The high brightness temperatures of the EVN-detected features indicate the AGN origin of the radio emission. As shown in the previous section, the brighter EVN-detected feature, A, is co-spatial with the e-MERLIN-detected component N. The flux density measured at $1.7$\, GHz with the e-MERLIN, $\sim 3.2$\,mJy, is slightly higher than the EVN value. This in principle can arise from variability or a resolution issue. The largest recoverable size, determined by the shortest baseline of the array (see Fig.\,\ref{fig:uvplot}), is $\sim120$\,mas for the EVN observation, smaller than the fitted size of component N in the e-MERLIN image ($\sim 152$\,mas). Thus, the lower flux density of component A is most likely caused by EVN resolving out some of the flux density contained in the extended feature and thus recovered by e-MERLIN. This indicates that the northern emitting region (component N) is extended but contains a mas-scale compact feature (component A).

While the EVN observation was conducted at a single frequency only, one can use the dual-frequency e-MERLIN measurements to infer a limit on the power-law radio spectral index of component A (Fig.~\ref{fig:spectralindex}). If we assume that all of the $5$-GHz flux density of component N is contained within a compact mas-scale feature, detectable with EVN at $5$\,GHz, it would imply a spectral index of $\alpha_\textrm{A} \approx -1.9$. This value can be regarded as an upper limit on the spectral index of the compact feature, since similarly to the lower frequency, some amount of the flux density may be resolved out. Such spectral index is often related to the lobe-like features seen in radio galaxies, while the cores of radio-loud AGN have usually flat or inverted radio spectra at cm wavelengths \citep[e.g.,][and references therein]{hovatta_spectralshape}.

The fainter feature in the EVN image, component B (Fig.~\ref{fig:EVN}) cannot be seen in the e-MERLIN images, neither at $1.5$\,GHz (Fig.~\ref{fig:eMERLIN-L}) nor at $5$\,GHz (Fig.~\ref{fig:eMERLIN-C}). Its non-detection at $5$\,GHz could naturally be explained by the component having a steep spectrum, $\alpha\lesssim -2.4$, rendering it too faint for the sensitivity of the $5$-GHz e-MERLIN observation. Such extremely steep radio spectra are seen in AGN remnants, old plasma no longer fed by an active jet. Alternatively, old plasma can be re-energized by turbulence or shocks in merging systems (radio phoenices), and show steeply rising radio spectrum towards lower frequencies \citep[e.g., ][and references therein]{Shulevski_2024}. However, its compact, few kpc-scale size is not compatible with the known radio phoenices extending for $\sim 100$\,kpc size \citep[e.g., ][]{Bruno_2025}. On the other hand, the estimation of the spectral index, its extreme negative value may be the result of surface brightness sensitivity and angular resolution effects.

Based on its brightness in the higher-resolution EVN image, component B should have been visible at a similar significance level as component S in the $1.5$-GHz e-MERLIN image. Nevertheless, the rms noise level of the image could have been estimated too optimistically. The $1\sigma$ rms noise level of the $1.5$-GHz e-MERLIN image at the approximate position of component B is $\sim 0.13$\,mJy\,beam$^{-1}$, about $3$ times larger than measured across the whole image, while the largest noise peak is $\sim 0.22$\,mJy\,beam$^{-1}$. Thus, it could be that component B was just below the sensitivity limit of the $1.5$-GHz e-MERLIN observation. Additionally, component B could also be variable. A flux density drop of $\sim 50\,\%$ within $\sim 1.2$\,months time in the rest frame of the source ($6$\,months elapsed between the EVN and e-MERLIN observations) is able to account for the non-detection of component B with the e-MERLIN at $1.5$\,GHz. If true, such variability is more characteristic for the radio core of the AGN. However, the radio core is not expected to have steep radio spectrum.

At the position of feature S, there is no sign of any EVN detection, most probably indicating the resolved nature of this component. Moreover, feature S also has a steep radio spectrum indicated by the e-MERLIN data (Fig.~\ref{fig:spectralindex}). Due to the faintness of this feature at $5$\,GHz, also indicated by the large relative error of its spectral index, it may have even steeper radio spectrum than the one formally obtained, $\alpha_\mathrm{S}\sim -1.7$. 

Thus, neither of the detected radio features can be clearly identified with the central core of the radio-emitting AGN that is expected to have flat or inverted synchrotron self-absorbed spectrum. The sizes and spectral indices of N and S are more compatible with extended jet-related emissions containing mas-scale compact hot spot such as feature A in the EVN observation. Due to the scarcity of radio data, no firm conclusion can be drawn concerning the nature of component B.

\subsection{The radio components and the {\it JWST}-detected features}
\label{sec:jwst}

The continuum-subtracted H$\alpha$ line emission, as obtained from the {\it JWST}/NIRSpec IFU observations, together with the results of the e-MERLIN radio observations are shown in Fig.\,\ref{fig:JWST-eMERLIN}. Several bright emission-line components lie along the general direction of the radio axis. The radio-emitting regions are close to two large ionized areas \citep[labeled L1 and L3 in][ and in Fig.\,\ref{fig:JWST-eMERLIN}]{Saxena_2026}, but they do not coincide with either of them. Based on the {\it JWST} observations, it is believed that the radio emission emerges from a bright optical continuum and emission-line source \citep[labeled as C2 in][and in Fig.\,\ref{fig:JWST-eMERLIN}]{Saxena_2026} located in between the two radio lobes. The mas-scale radio component, B, is within $\sim 0.1''$ of this continuum object. C2 remained unresolved in the NIRCam images, thus its extent can only be approximated as $\sim 0.1''-0.15''$, corresponding to a linear size of $\sim (0.7-1)$\,kpc.

The {\it JWST} data furthermore show evidence for perturbed gas, especially close to the location of the southern hotspot. Besides these jet--gas interactions, other sources likely contributing to the strong alignment observed are the ionization by a central, obscured active nucleus and an infalling low-mass starburst galaxy at the southern edge of the field \citep[object C6 in][]{Saxena_2026}.

\subsection{Radio emission at larger scales} \label{sec:large_scale}

According to the $1.4$-GHz NRAO VLA Sky Survey \citep{nvss}, J1530$+$1049 has a flux density of $(7.4\pm0.5)$\,mJy. At the same frequency but with higher resolution, the Faint Images of the Radio Sky at Twenty-centimeters (FIRST) survey measured $(7.25\pm0.15)$\,mJy \citep{first_helfand}. \cite{Saxena_sample} reported a similar flux density value, $(7.5\pm0.1)$\,mJy, measured in a higher-resolution VLA A-configuration observation. More recently, J1530$+$1049 was detected in the the Rapid Australian Square Kilometre Array Pathfinder (ASKAP) Continuum Survey (RACS with flux densities of $(6.4 \pm 0.6)$\,mJy, and $(5.6 \pm 0.8)$\,mJy at $1.37$\,GHz, and $1.66$\,GHz, respectively \citep{racs-mid, racs-high}. All these values exceed the sum of flux densities of components N and S detected with e-MERLIN at close frequencies. However, no self-calibration was performed on the e-MERLIN data, therefore coherence loss can result in lower measured flux density values. 

J1530$+$1049 was detected at $3$\,GHz in the three epochs of the VLA Sky Survey \citep[VLASS, ][]{vlass_lacy}. We downloaded the three images from the Canadian Astronomy Data Centre (CADC\footnote{\url{www.cadc-ccda.hia-iha. nrc-cnrc.gc.ca/en/vlass/}, accessed on 12 Sep. 2025}), and used \texttt{AIPS} to fit them with a single Gaussian. The flux densities agree within the error, their average value is $(2.1\pm0.5)$\,mJy. This exceeds the flux density expected from the spectral indices derived from the e-MERLIN observations (Fig.~\ref{fig:spectralindex}), $\sim 1$\,mJy. While a frequency-independent coherence loss of the e-MERLIN data will not alter the obtained spectral index, it can underestimate the predicted $3$-GHz flux density. 

Below $1$\,GHz, J1530$+$1049 is brighter, as indicated by the lowest available frequency measurement of the RACS, $888$\,MHz \citep{rapid_askap}, $(17.1\pm0.71)$\,mJy, and by the TIFR Giant Meterwave Radio Telescope Sky Survey \citep[TGSS, ][]{TGSS} at $150$\,MHz, $(170\pm19)$\,mJy (Fig.~\ref{fig:spectralindex}). These low-resolution flux density values imply a spectral index of $\alpha_{0.15\mathrm{\,GHz}}^{3\mathrm{\,GHz}}=-1.5\pm0.1$, but also indicate a flattening of the radio spectrum between $888$\,MHz and $150$\,MHz, corresponding to $\sim5$\,GHz and $750$\,MHz, respectively, in the rest frame of the source at $z=4$.

The $1.4$-GHz radio power of J1530$+$1049 is between $\sim(3-4) \times 10^{27} \mathrm{\,W\,Hz}^{-1}$, depending on whether the highest $1.4$-GHz flux density and the broad-band spectral index values are used, or whether the flux density and spectral index are taken from the e-MERLIN observations. The linear size of the source is $\sim 5.5$\,kpc. It is an order of magnitude fainter and smaller version of the radio galaxy J1420$+$1205 located at similar redshift ($z=4.026$) and also showing several steep-spectrum features at $100$-mas scale \citep{J1420}. 

The linear size and radio power of J1530$+$1049 are similar to those of the medium-sized symmetric objects (MSOs) studied by \cite{Fanti_2001} out to a redshift of $\sim 4$. In Fig.\,3 in \cite{Orienti_2025}\footnote{The unit of the horizontal axis is erroneously given in pc, it is in kpc.}, J1530$+$1049 could be placed on the evolutionary track shown for objects with jet power of $10^{46}$\,erg\,s$^{-1}$. MSOs are regarded as young radio sources eventually evolving into large-scale radio galaxies \citep[e.g.,][]{Magda2010, anbaan, Orienti_2025}. Alternatively, their jets can be frustrated, confined within their host galaxies, unable to break through the environment and grow into larger structures. According to \cite{Orienti_2025}, the progenitor objects of the high-power MSOs are less likely to be influenced by jet instabilities, and expected to eventually evolve into Fanaroff--Riley type II radio galaxies.

Recently, \cite{Stanghellini_2025} studied the high-resolution radio morphology and the optical structure of the host galaxies in radio sources containing bright, sub-galactic sized radio jets. They distinguished two groups, those with aligned radio structures likely evolving into radio galaxies in the future and the ones with more complex radio morphologies. They suggested the latter objects reside in merging systems, where the jet interaction with the disturbed interstellar medium, or orbital motion of the jet-emitter AGN in a binary system give rise to rapidly re-orienting jets. There is no clear indication of large bending structures in the radio morphology of J1530$+$1049. Nevertheless, the three radio-emitting regions, S, B, N (with the hotspot A) are not positioned along a straight line. Additionally, the \textit{JWST} data revealed a quite disturbed environment around the radio structure, showing the dense region of merging galaxies. Thus, the environment of J1530$+$1049 may influence its radio structure similarly as found by \cite{Stanghellini_2025}. More sensitive, high-resolution radio observations conducted at lower frequencies, allowing for the detection of the steep-spectrum jet-related emission, can reveal possible jet disturbances.

\section{Summary}
\label{sum}

We imaged the radio emission of a radio galaxy at $z=4$, J1530$+$1049, both at mas-scale and $100$-mas scale angular resolutions. We detected two faint, mas-scale compact features at $\sim 2.8$\,kpc projected separation with the EVN at $1.7$\,GHz. Their high brightness temperature values clearly indicate AGN-related origin.

The northern one could be detected as part of a larger component in our e-MERLIN observation at $1.5$\,GHz and $5$\,GHz. At larger scales, we detected another steep-spectrum, extended component at $\sim 5.5$\,kpc separation from the brightest feature in both e-MERLIN observations. According to the flux densities of the radio features, all have steep power-law spectra. Thus neither of them can be identified as the centre of the AGN. All these features are most probably related to the lobes or hot spots in the AGN, possibly also indicating places of interaction with the surrounding medium. Among the detected radio features, the faintest compact component (B), only detected at the $1.7$-GHz EVN observation, is located close to the optical continuum source identified with the AGN. It might be related to the AGN core, however, more sensitive observations are needed to ascertain its nature.

The radio power and linear extent of the object place it among the medium-sized symmetric objects which are smaller and presumably younger versions of radio galaxies. A comparison with recent observations with the {\it JWST} reveals that the compact radio source is embedded in a very dense region of merging galaxies. Bright knots of emission-line gas are strongly aligned with the radio axis, but are distributed over a total extent about five times larger than the size of the radio source. These data point to a picture of a massive, forming galaxy that hosts at least one active supermassive black hole interacting with the surrounding gas through its (compact) radio jets.

\begin{acknowledgements}
We thank the anonymous referee for their comments which helped us improve our paper.
      The European VLBI Network is a joint facility of independent European, African, Asian, and North American radio astronomy institutes. Scientific results from data presented in this publication are derived from the following EVN project code(s): RSG11. The e-VLBI research infrastructure in Europe was supported by the European Union’s Seventh Framework Programme (FP7/2007-2013) under grant agreement number RI-261525 NEXPReS. The research leading to these results has received funding from the European Commission Horizon 2020 Research and Innovation Programme under grant agreement No. 730562 (RadioNet).
      e-MERLIN is a National Facility operated by the University of Manchester at Jodrell Bank Observatory on behalf of STFC. Scientific results from data presented in this publication are derived from the e-MERLIN project CY8205.
      On behalf of the ``Interferometric studies of radio-loud active galactic nuclei'' project, we are grateful for the possibility to use the HUN-REN Cloud \citep[see][https://science-cloud.hu/]{Heder2022} which helped us achieve the results published in this paper. This project was supported by the HUN-REN Hungarian Research Network. K\'EG and SF received funding from the Hungarian National Research, Development and Innovation Office (NKFIH excellence grant TKP2021-NKTA-64). MVM research has been funded by grant Nr. PID2021-124665NB-I00  by the Spanish Ministry of Science and Innovation/State Agency of Research MCIN/AEI/10.13039/501100011033 and by "ERDF A way of making Europe".
\end{acknowledgements}

\bibliography{ref}

\end{document}